\documentclass[prl,twocolumn,twoside,superscriptaddress,showpacs,floatfix]{revtex4}
\usepackage{dcolumn}
\usepackage{amsmath}
\usepackage{graphicx}

\newlength{\figurewidth}
\newcommand{\PastSeq}    { \stackrel{\leftarrow}{\omega} }
\newcommand{\FutSeq}    { \stackrel{\rightarrow}{\omega} }

\begin{document}

\title{Discovering Planar Disorder in Close-Packed Structures from
X-Ray Diffraction:\\ Beyond the Fault Model}
\author{D. P. Varn}
\affiliation{Santa Fe Institute, 1399 Hyde Park Road, Santa Fe, NM 87501}
\affiliation{Department of Physics and Astronomy, University of Tennessee,
Knoxville, TN 37996}
\author{G. S. Canright}
\affiliation{Department of Physics and Astronomy, University of Tennessee,
Knoxville, TN 37996}
\affiliation{Telenor Research and Development, 1331 Fornebu, Norway}
\author{J. P. Crutchfield}
\affiliation{Santa Fe Institute, 1399 Hyde Park Road, Santa Fe, NM 87501}

\date{February 28, 2002}

\begin{abstract}
We solve a longstanding problem---determining structural information for
disordered materials from their diffraction spectra---for the case of planar
disorder in close-packed structures (CPSs). Our solution offers the most
complete possible statistical description of the disorder, and, from it, we
find the minimum effective memory length of disordered stacking sequences.
We also compare our model of disorder with the so-called fault model (FM)
and demonstrate that in simple cases our approach reduces to the FM, but in
cases that are more complex it provides a general and more accurate structural
description than the FM. We demonstrate our technique on two previously
published zinc sulphide diffraction spectra.
\end{abstract}

\pacs{
  61.72.Dd,   %  Experimental determination of defects by diff/scattering
  61.10.Nz,   %  Single crystal/powder diffraction
  61.43.-j,   %  Disordered Solids
  81.30.Hd    %  Constant-Composition Solid State Transformations
  ~~~~~~~~~~~~Santa Fe Institute Working Paper 02-03-014
  }

\maketitle

Describing the structure of solids---by which we simply mean the placement of
atoms in (say) a crystal---is essential to a detailed understanding of
material properties. Crystallography has long used the sharp Bragg peaks
in X-ray diffraction spectra to infer crystal structure. For those cases where
there is diffuse scattering, however, finding---let alone describing---the
structure of a solid has been more difficult~\cite{W85}. Indeed, it is known
that without the assumption of crystallinity, the inference problem has no
unique solution~\cite{G63}. Moreover, diffuse scattering implies that a solid's
structure deviates from strict crystallinity. Such deviations can come in
many forms---Schottky defects, substitution impurities, line dislocations,
and planar disorder, to name a few. Of these, here we consider only planar
disorder; that caused by one plane of atoms slipping relative to another in
a layered material. This kind of disorder is known to be prevalent in a
broad class of materials called {\em polytypes}.

First discovered in SiC by Baumhauer~\cite{B12} in 1912, polytypes
~\cite{T91,SK94} are solids built up from identical layers, called
{\em modular layers} (MLs)~\cite{VC01} that differ only in their stacking
orientation. A polytype is simply described by its {\em stacking
sequence}---the one-dimensional list of successive orientations found as
one moves along the stacking direction. We refer to the effective stochastic
process induced by scanning the list as the {\rm stacking process}. In the
H\"{a}gg notation~\cite{SK94} for stacking sequences one replaces the set
$\{A, B, C\}$ of allowed orientations with a binary alphabet
$\mathcal{A} = \{ 0,1 \}$: an ML is labeled `1', if it is cyclically related
to the preceding ML, or $0$, if it is not.

Polytypism is found in dozens of materials; one of the best studied is ZnS.
There are approximately $185$ identified crystalline structures~\cite{T91} and
many samples exhibit varying degrees of disorder. Notably, some ZnS crystals
have unit cells extending over $100$ MLs~\cite{SK94}.
These different stacking sequences can occur under virtually identical
thermodynamic growth conditions. The mystery of polytypism then is two-fold:
How can so many different structures (crystalline and noncrystalline) exist?
And, what are the source and range of interlayer interactions that produce
these structures? Over the last fifty years, considerable effort has been
expended to understand polytypism, with over a dozen theories having been
proposed; but a general explanation is still lacking~\cite{T91,SK94}. 

Attempts to describe planar disorder in CPSs have a long history.
Early studies~\cite{HT42,W42} focused on stacking {\em errors} or {\em faults}
that permeated a parent crystal. Different kinds of stacking faults were
postulated, such as {\em growth faults}, {\em deformation faults}, and
{\em layer displacement faults}~\cite{SK94}. In this {\em fault model} (FM)
theory,
stacking faults were introduced randomly into the parent crystal and their
effect on the intensity, placement, and broadening of Bragg peaks was
calculated as a function of the fault frequency. These efforts met with good
success for several weakly faulted specimens as such cobalt~\cite{W42} and
lithium~\cite{BW86}. However, for polytypes such as ZnS and SiC, the random
insertion of faults often did not describe the observed Bragg peaks well.
More sophisticated models~\cite{SK94,G00,G01} were introduced which attempted
to account for nonrandom fault insertion by assuming the existence of some
``coordination'' between faults. These more complicated models gave mixed
results.

We find several drawbacks to the FM. The first is the need to assume a single
parent crystalline structure into which stacking defects are introduced; this
precludes the description of disorder interspersed between distinct crystal
structures. In some polytypes, such as ZnS, for example, there is considerable
interest in characterizing the transformation between the hexagonally
closed-packed (HCP) structure and the cubic close-packed (CCP) structure when
the crystal is subjected to an external stress, such as annealing~\cite{SK94}.
In these cases, there is no single parent crystal in which  to introduce
faulting. A second drawback is not inherit to the FM, but in the way it is
analyzed~\cite{SK94}. By considering only the effects of faulting on Bragg
peaks, information in the diffuse scattering is ignored. (Our second example
below demonstrates how misleading this can be.) Our final difficulty is with
the FM generally. Here we show that it is not possible to uniquely identify
and assign faulting sequences to disordered crystals, except in those
special cases to which the FM is restricted.

In this Letter, we introduce a novel method for discovering and describing
disordered stacking sequences in CPSs that overcomes these difficulties. 
We analyze two previously published diffraction spectra for polytypic
ZnS~\cite{SK94} and compare our results with those of the FM. 

Our method can be broken into three parts. In the first, we use a diffraction
spectrum to find average correlations between MLs as a function of
the number $n$ of separating layers. If we assume that the MLs themselves
are undefected, that each ML has the same scattering power, and that the spacing
between MLs is independent of the local stacking arrangement, then correlation
factors (CFs), $Q_c(n)$ and $Q_a(n)$~\cite{YC96}, can be found by Fourier
analysis of the diffraction spectrum~\cite{G63}. $Q_c(n)$ and $Q_a(n)$ are
defined as the probability that any two MLs at separation $n$ are cyclically
or anticyclically related, respectively.

In the second part of our approach, we infer the spatial patterns of MLs that
reproduce these CFs by reconstructing an $\epsilon$-machine~\cite{CMech},
which describes the minimal effective states of the stacking process. Assume
we know the probability $p(\omega)$ of stacking sequences $\omega$. At each
ML in a stacking sequence define the ``past'' $\PastSeq$ as those MLs already
seen and the ``future'' $\FutSeq$ as those yet to be seen:
$\omega = \PastSeq \FutSeq$. The effective states of the stacking process then
are defined as the {\em sets} of pasts $\PastSeq$ that lead to statistically
equivalent futures:
\begin{equation}
\PastSeq_i \sim \PastSeq_j {\rm~if~and~only~if~}
  p(\FutSeq | \PastSeq_i) = p(\FutSeq | \PastSeq_j) ~.
\label{CausalStateDef}
\end{equation}
These equivalence classes of pasts are the stacking process's {\em causal
states}. Along with their transitions, they comprise the process's
{\em $\epsilon$-machine}---a statistical description of the ensemble
of spatial patterns that produces the stacking distribution $p(\omega)$.
It has been shown that the $\epsilon$-machine is the optimal predictor of
minimal size (as measured by the number of states) of a process, and, up to
state-relabeling, it is the unique such description~\cite{CMech}.

To find the causal states we must first estimate the probability $p(\omega)$
of stacking sequences $\omega$ averaged over the sample. Note that, from
conservation of probability, $p(u) = p(0u) + p(1u) = p(u0) + p(u1)$, for all
$u \in {\mathcal{A}}^r$, where $\mathcal{A}^r$ is the set of all sequences
of length $r$. Additionally, the probabilities for sequences of the same
length are normalized: $\sum_{\omega \in \mathcal{A}^{r+1}}p(\omega) = 1$.
Together these constraints provide $2^r$ independent relations among
probabilities for the $2^{r+1}$ possible stacking sequences of length $r+1$.   

The other $2^r$ constraints come from relating CFs to sequence probabilities 
via
\begin{eqnarray}
Q_{\alpha}(n) = \sum_{\omega \in \mathcal{A}^n_\alpha} p(\omega) ~,
\label{eq:Q_p}
\end{eqnarray}
where $\mathcal{A}^n_\alpha$ is that subset of length-$n$ sequences with a
cyclic ($\alpha = c$) or an anticyclic ($\alpha = a$) rotation between MLs
at separation $n$. We take as many of these latter relations as necessary to
form a complete set of equations. At a fixed $r$, the set of equations describes
the stacking sequence as generated by an $r^{\textrm{th}}$-order Markov process.
At $r = 3$ one encounters the first nonlinearities due to the necessity of
using CFs at $n = 5$ to obtain a complete set of equations. We rewrite the
probability of sequences of length $n = 5$ in terms of the conditional
probabilities of those at $n = 4$, and it is this mapping that is nonlinear.
We solve numerically for the stacking sequence
probabilities $p(\omega)$ and then find the set of causal states using the
equivalence relation Eq. (\ref{CausalStateDef}). The causal-state transitions
are estimated from the conditional distributions of the next ML orientation
given pasts $\PastSeq$ associated with each causal state.

In the third and final part, we begin with the $r = 1$ reconstructed
$\epsilon$-machine, use it to generate a sample stacking sequence (here we
used length $400,000$), and from this we estimate the $\epsilon$-machine's
predicted CFs and diffraction spectrum. We then compare the latter to the
experimental diffraction spectrum. If there is not sufficient agreement,
we increment $r$ and repeat the reconstruction and comparison. The resulting
$r$ is called the stacking process's {\it memory length}, since it is the
amount of history (in MLs) one must use to optimally predict the process.

ZnS can be thought to have a CPS with a basis composed of two atoms, zinc and 
sulphur, with the sulphurs displaced one quarter of a body diagonal (as referred
to the conventional unit cell) along the stacking direction~\cite{SK94}. We
take an ML to be this zinc sulphur pair arranged in a hexagonal net~\cite{VC01},
giving (as with any CPS) three absolute orientations for the MLs, but only two
relative orientations for neighboring layers. We correct the experimentally
obtained diffraction spectrum for the atomic scattering factors, the structure
factor, dispersion factors, and polarization of the incident
radiation~\cite{HW92}. 

\begin{figure}
\begin{center}
\resizebox{!}{5cm}{\input{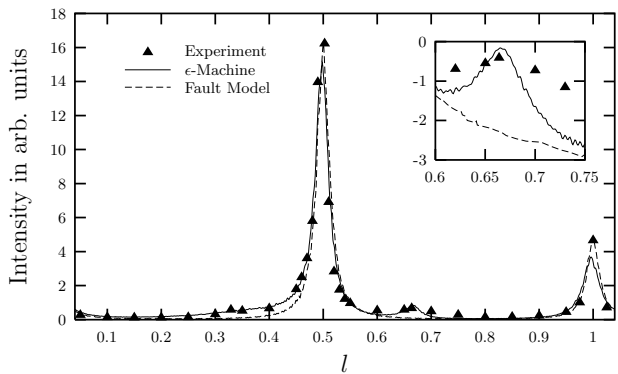}}
\end{center}
\caption{Comparison of the experimental diffraction spectrum SK134 along the
  $10.l$ row (triangles) for a disordered ZnS single crystal~\cite[p. 134]{SK94}
  with spectra estimated from the FM with $5\%$ deformation faulting (dashed
  line) and $r = 3$ $\epsilon$-machine (solid line). The vertical scale in the
  inset is logarithmic intensity.
  }
\label{fig:dpSK134a.r.3}
\end{figure}

We now give the results for $\epsilon$-machine reconstruction for two
experimental diffraction spectra, SK134 and SK135 from~\cite{SK94}. Let $l$
be a continuous variable that indexes the magnitude of the perpendicular
component of the diffracted wave $k = 2\pi l/c$, where $c$ is the spacing
between adjacent MLs. We select a unit interval in $l$ on which to analyze
each spectra. Since many diffraction spectra suffer from experimental
error~\cite{SK94}, we show elsewhere~\cite{VC02} that there are relations
that the CFs must obey for any CPS and that we can use them to select 
a relatively error-free $l$-interval. The spectra from experiment and
$\epsilon$-machine reconstruction are normalized.

\begin{figure}  
\begin{center}
\resizebox{!}{4.5cm}{\includegraphics{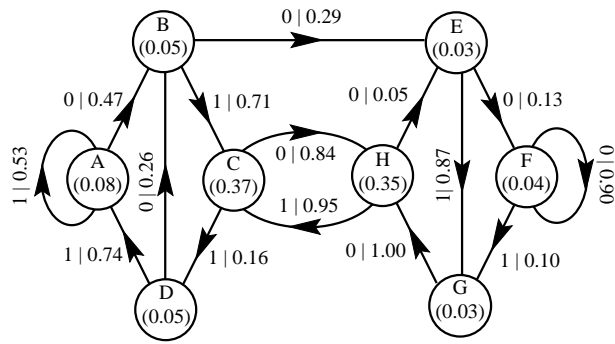}}
\end{center}
\caption{The recurrent causal states $\{ {\rm A}-{\rm H} \}$ of the
  reconstructed
  $\epsilon$-machine estimated from the experimental diffraction spectrum
  SK134 of Fig. \ref{fig:dpSK134a.r.3} with $r = 3$. Asymptotic state
  probabilities are given in parentheses; edge label $s|p$ indicates a
  transition on symbol $s$ with probability $p$.
  }
\label{SK134.db3}
\end{figure}

The triangles in Fig.~\ref{fig:dpSK134a.r.3} show the experimental diffraction
spectrum SK134 along the $10.l$ row for an HCP ZnS crystal annealed at 300 C
for one hour. Sebastian and Krishna~\cite{SK94} attribute the observed disorder
to the introduction of $5\%$ deformation faulting. (This is the FM predicted
spectrum given as a dashed line in Fig.~\ref{fig:dpSK134a.r.3}). We find that
the smallest-$r$ $\epsilon$-machine that gives adequate agreement (solid
line) with experiment is estimated at $r = 3$; it is shown in
Fig.~\ref{SK134.db3}.

It is possible to give an approximate equivalent of this $\epsilon$-machine in
terms of the FM, but we stress that this decomposition is {\it not unique}.
We associate each closed, nonintersecting loop (called a {\em simple cycle} or
SC~\cite{CW96}) in the $\epsilon$-machine with either a crystal structure or
a fault. In this way, $\epsilon$-machines directly describe familiar structures
in polytypes. For instance, the closed loop between causal states $\rm C$ and
$\rm H$ in Fig.~\ref{SK134.db3} implies a stacking sequence
$\ldots 010101 \ldots$,
which is simply the H\"{a}gg notation for the HCP structure. One concludes,
then, that there is no qualitative difference between what one calls faults
and crystal structure. The distinction is, in fact, quantitative and one of
convenience---crystal structures have relatively high probabilities, as
opposed to the rarer faults. For the most general $r = 3$ $\epsilon$-machine,
it is known that there are $19$ such SCs~\cite{T90}. Since eight independent
CFs are sufficient to specify an $r=3$ $\epsilon$-machine, the problem of
decomposing the $\epsilon$-machine into SCs is underdetermined. This conclusion
holds for all $r \geq 2$. Therefore, without a fortuitous vanishing of causal
states or transitions, the fault description is not unique.  

For the sake of comparison with previous FM analyses, we decompose the
$\epsilon$-machine in Fig.~\ref{SK134.db3} into SCs with the assumption that
faults corresponding to SCs of length $7$ or greater are not present. We can
then assign an {\it approximate} fault distribution for SK134 (second column)
as follows and compare it to that of ~\cite{SK94} (third column): 
\begin{center}
\( \begin{array}{lrr}
   \textrm{HCP}                      &  $ 64\% $  & $95\%$ \\
   \textrm{CCP}                      &  $  8\% $  & $ 0\%$ \\
   \textrm{Deformation Fault}        &  $ 16\% $  & $ 5\%$ \\
   \textrm{Growth Fault}             &  $  6\% $  & $ 0\%$ \\
   \textrm{Layer Displacement Fault} &  $  6\% $  & $ 0\%$
\end{array} \)
\end{center}  
The $\epsilon$-machine description of the crystal differs significantly from
that of Sebastian and Krishna~\cite{SK94}. While we both find qualitatively
that deformation faulting is important, we also detect CCP structures, as well
as growth faults and layer displacement faults. Overall, $\epsilon$-machine
analysis finds a much more disordered crystal. This is borne out when comparing
the FM and $\epsilon$-machine diffraction spectra. Fig. \ref{fig:dpSK134a.r.3}
shows that, while both agree reasonably well with experiment at the broadened
peaks at $l = 0.5$ and $1$, the $\epsilon$-machine is in better agreement along
the shoulders of the Bragg peaks, as well as at the rise in broadband intensity
at $l \approx 0.67$ (inset in Fig. \ref{fig:dpSK134a.r.3}).

\begin{figure}
\begin{center}
\resizebox{!}{5cm}{\input{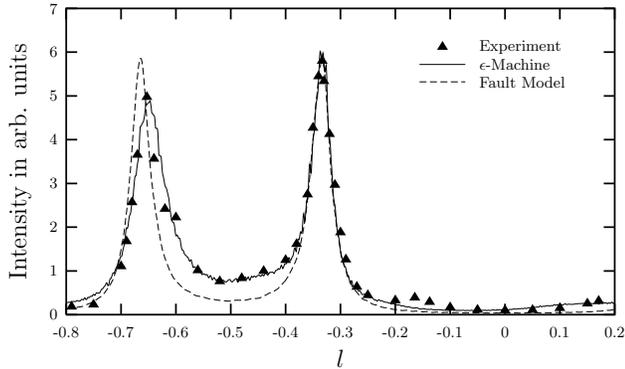}}
\end{center}
\caption{Comparison of the experimental diffraction spectrum SK135 along the
  $10.l$ row (triangles) for a disordered CCP ZnS single
  crystal~\cite[p. 135]{SK94} with the diffraction spectra calculated from the
  the FM with $12\%$ twined faulting (dashed line) and $r = 3$
  $\epsilon$-machine (solid line).
  }
\label{fig:dpSK135a.r.3}
\end{figure}

\begin{figure}[b]
\begin{center}
\resizebox{!}{4.5cm}{\includegraphics{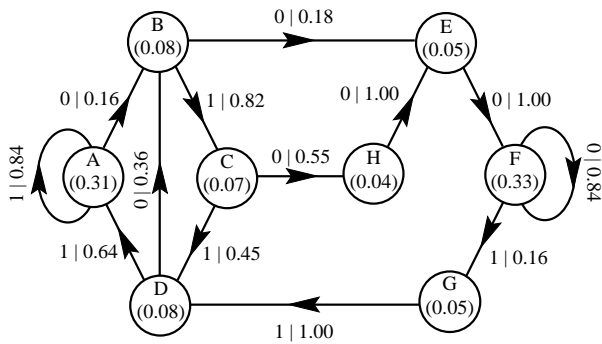}}
\end{center}
\caption{Recurrent states of the reconstructed $\epsilon$-machine for the
  experimental diffraction data SK135 of Fig. \ref{fig:dpSK135a.r.3}
  using $r = 3$.
  }
\label{SK135.db3}
\end{figure}

Fig.~\ref{fig:dpSK135a.r.3} plots the experimental diffraction spectrum along
the $10.l$ row (triangles) for a HCP ZnS crystal  annealed at 500 C for one
hour. Sebastian and Krishna~\cite{SK94} calculate a twin-fault probability
of $12\%$ from the observed half-widths of the peaks. The calculated diffraction
spectrum for such a faulting mechanism is shown in Fig.~\ref{fig:dpSK135a.r.3}
(dashed line). Only the peak at $l = -0.33$ was used to find the faulting
mechanism, and one sees that the FM reproduces it well. However, the second
peak at $l = -0.67$ is poorly represented, as is the diffuse scattering between
the two peaks. This demonstrates the pitfalls in simply fitting an FM to a
single Bragg peak, ignoring the information contained in other peaks and
in the diffuse scattering.
We also note that the small rise in diffracted intensity at $l \approx -0.16$
is likewise missed by the FM. The $\epsilon$-machine spectrum (solid line)
also misses this rise, but otherwise is in excellent agreement with the
experiment. Fig.~\ref{SK135.db3} shows the reconstructed $\epsilon$-machine
obtained at $r=3$. The large probabilities for causal states $\rm A$ and
$\rm F$ and their large self-loop transition probabilities, associated with
stacking sequences $\ldots 1111 \ldots$ and $\ldots 0000 \ldots$, indicate
that this is a twined-CCP crystal. The missing ${\rm H} \rightarrow {\rm C}$
causal-state transition, and so the resulting absence of the
$\ldots 0101 \ldots$ stacking, implies that the original HCP structure
has been eliminated.  

In conclusion, we presented a novel method for discovering and
describing planar disorder and organization in CPSs. We demonstrated that
the FM, both in conception and practice, is unable to accommodate the variety
of stacking arrangements possible in nature. In contrast, $\epsilon$-machines
provide a unique description of structure and can be used to quantify any
amount of disorder. We argued that simply examining the effects of disordered
stacking on the Bragg peaks is insufficient to properly detect the disorder
present. Moreover, we quantified the memory length for disordered 1D systems;
for the ZnS samples considered it was 3 MLs. Thus, the origins of polytypic
ordering lie outside the physical 1D Hamiltonian, from which one calculates
a physical interaction range of 1 ML ~\cite{EN90}. 

While we performed $\epsilon$-machine reconstruction only to $r=3$, the
extension to higher $r$ is straightforward, but quite demanding computationally.
Additionally, knowing the $\epsilon$-machine for a material's spatial patterns
allows calculation of physically relevant quantities. We show elsewhere that
given the coupling constants between MLs~\cite{EN90}, we can determine the
average stacking-fault energy for a disordered crystal~\cite{VC02}. We expect
that other physical parameters will be amenable to calculation directly from
$\epsilon$-machines. 

The authors thank David Feldman for comments on the manuscript.
This work was supported at the Santa Fe Institute under the Network Dynamics
Program funded by Intel Corporation and under the Computation, Dynamics, and
Inference Program via SFI's core grants from the National Science and
MacArthur Foundations. Direct support was provided by NSF grants DMR-9820816
and PHY-9910217 and DARPA Agreement F30602-00-2-0583. DPV's visit to SFI
was supported by the NSF Physics Graduate Fellows Program.


\begin{thebibliography}{99}

\bibitem{W85}
T. R. Welberry,
Rep. Prog. Phys.
{\bf 48},
1543
(1985).

\bibitem{G63}
A. Guinier,
{\it X-Ray Diffraction in Crystals, Imperfect Crystals and Amorphous Bodies.}
(W.H. Freeman and Company, San Francisco 1963). 

\bibitem{B12}
H. Baumhauer, 
Z. Krist. 
{\bf 50}, 
33 
(1912).  

\bibitem{T91}
G. C. Trigunayat,
Solid State Ionics.
{\bf 48},
3
(1991).

\bibitem{SK94}
M. T. Sebastian and P. Krishna,
{\it Random, Non-Random and Periodic Faulting in Crystals.}
(Gordon and Breach Science Publishers, Langhorne, Pennsylvania 1994).

\bibitem{VC01}
D. P. Varn and G. S. Canright,
Acta Cryst. A
{\bf 57},
4
(2001).

\bibitem{HT42}
S. Hendricks and E. Teller,
J. Chem. Phys.
{\bf 10},
147
(1942).

\bibitem{W42}
A. J. C. Wilson,
Proc. Roy. Soc.
{\bf A180},
277
(1942).

\bibitem{BW86}
R. Berliner and S. A. Werner,
Phys. Rev. B
{\bf 34},
3586
(1986).

\bibitem{G00}
J. B. Gosk,
Crys. Res. Tech.
{\bf 35},
101
(2000).

\bibitem{G01}
J. B. Gosk,
Crys. Res. Tech.
{\bf 36},
197
(2001).

\bibitem{YC96}
J. Yi and G. S. Canright,
Phys. Rev. B
{\bf 53},
5198
(1996).

\bibitem{CMech}
J. P. Crutchfield and K. Young,
Phys. Rev. Lett. 
{\bf 63}, 105
(1989);
J. P. Crutchfield and D. P. Feldman,
Phys. Rev. {\bf E}
{\bf 55}, R1239
(1997); and
C. R. Shalizi and J. P. Crutchfield,
J. Stat. Phys.
{\bf 104}, 819
(2001).

%\bibitem{CY89}
%J. P. Crutchfield and K. Young,
%Phys. Rev. Lett. 
%{\bf 63},
%105
%(1989).
%
%\bibitem{CF97}
%J. P. Crutchfield and D. P. Feldman,
%Phys. Rev. {\bf E}
%{\bf 55},
%R1239
%(1997).
%
%\bibitem{SC01}
%C. R. Shalizi and J. P. Crutchfield,
%J. Stat. Phys.
%{\bf 104},
%819
%(2001).
%
%\bibitem{C94}
%J. P. Crutchfield,
%Physica D
%{\bf 75},
%11
%(1994).

\bibitem{HW92}
T. Hahn, A. J. C. Wilson and U. Shmueli,
{\it International Tables for Crystallography, $3^{\mathrm{rd}}$ revised edition}.
(Kluwer Academic Publishers, Boston 1992).

\bibitem{VC02}
D. P. Varn, G. S. Canright and J. P. Crutchfield,
in preparation.
D. P. Varn. 
{\it Language Extraction from ZnS.}
Ph.D. Thesis, Univ. of Tenn., Knoxville
(2001).

\bibitem{CW96}
G. Canright and G. Watson,
J. Stat. Phys.
{\bf 84},
1095
(1996).

\bibitem{T90}
M. Teubner,
Physica A
{\bf 169},
407
(1990).

\bibitem{EN90}
G. E. Engel and R. J. Needs,
J. Phys. Cond. Mat.
{\bf 2},
367
(1990).

\end{thebibliography}
\end{document}